\documentclass[aps,prd,preprint,superscriptaddress,showpacs]{revtex4}
\usepackage{epsfig}
\usepackage{dcolumn}
\usepackage{bm}
\usepackage{amsmath}
\usepackage{amsfonts}
\usepackage{amssymb}
\def\nnd{\end{document}}

\def\nn{\nonumber}

\def\be{\begin{equation}}
\def\ee{\end{equation}}

\newcommand{\bea}{\begin{eqnarray}}
\newcommand{\eea}{\end{eqnarray}}
\newcommand{\bwt}{\begin{widetext}}
\newcommand{\ewt}{\end{widetext}}

\def\u
\def\hZ{\widehat Z}

\def\eed{\end{document}}

\def\m_Z{m_{\textrm {Z}}}

\renewcommand{\u}{\rm{u}}

\newcommand{\s}{{\rm{s}}}

\newcommand{\beq}{\begin{equation}}
\newcommand{\eeq}{\end{equation}}
\newcommand{\bq}{\begin{equation}}
\newcommand{\eq}{\end{equation}}
\newcommand{\ba}{\begin{array}}
\newcommand{\ea}{\end{array}}
\newcommand{\beqa}{\begin{eqnarray}}
\newcommand{\eeqa}{\end{eqnarray}}

\def\be{\beta}

\def\rm#1{\textrm{#1}}

\makeatother
\begin{document}
\title{Constraints on Neutrino Velocities Revisited}

\author{Yunjie Huo}

\affiliation{State Key Laboratory of Theoretical Physics,
Institute of Theoretical Physics, Chinese Academy of Sciences,
Beijing 100190, P. R. China}

\author{Tianjun Li}

\affiliation{State Key Laboratory of Theoretical Physics,
Institute of Theoretical Physics, Chinese Academy of Sciences,
Beijing 100190, P. R. China}

\affiliation{George P. and Cynthia W. Mitchell Institute for
Fundamental Physics and Astronomy, Texas A$\&$M University,
College Station, TX 77843, USA}

\author{Yi Liao}

\affiliation{Center for High Energy Physics, Peking University,
Beijing 100871, P. R. China}

\affiliation{School of Physics, Nankai University, Tianjin 300071, P. R. China}

\author{Dimitri V. Nanopoulos}

\affiliation{George P. and Cynthia W. Mitchell Institute for
Fundamental Physics and Astronomy, Texas A$\&$M University,
College Station, TX 77843, USA}

\affiliation{Astroparticle Physics Group,
Houston Advanced Research Center (HARC),
Mitchell Campus, Woodlands, TX 77381, USA}

\affiliation{Academy of Athens, Division of Natural Sciences, 28
Panepistimiou Avenue, Athens 10679, Greece}

\author{Yonghui Qi}

\affiliation{State Key Laboratory of Theoretical Physics,
Institute of Theoretical Physics, Chinese Academy of Sciences,
Beijing 100190, P. R. China}

\begin{abstract}

With a minimally modified dispersion relation for neutrinos, we
reconsider the constraints on superluminal neutrino velocities from
bremsstrahlung effects in the laboratory frame. Employing both the
direct calculation approach and the virtual $Z$-boson approach, we
obtain the generic decay width and energy loss rate of a
superluminal neutrino with general energy. The Cohen-Glashow's
analytical results for neutrinos with a relatively low energy are
confirmed in both approaches. We employ the survival probability
instead of the terminal energy to assess whether a neutrino with a
given energy is observable or not in the OPERA experiment. Moreover,
using our general results we perform systematical analyses on the
constraints arising from the Super-Kamiokande and IceCube
experiments.

\end{abstract}


\preprint{ACT-21-11, MIFPA-11-54}

\maketitle

\renewcommand{\thefootnote}{\arabic{footnote}} \setcounter{footnote}{0}%

\section{Introduction}

The OPERA collaboration has measured the velocities of muon
neutrinos ($\nu_{\mu}$) with mean energy about 17.5 GeV, which
travel from CERN to the Gran Sasso. It was found that the muon
neutrinos travel at a speed faster than light (so called propagating
superluminally), and the relative difference of the neutrino speed
$v_\nu$ with respective to that of light $c$ in the vacuum was
measured to be
\begin{eqnarray}
\delta' v_\nu &\equiv&  {{v_{\nu} -c}\over c} ~=~
\left( 2.37 \pm 0.32
~\textrm{(stat.)}^{+0.34}_{-0.24} ~\textrm{(sys.)} \right) \times
10^{-5}~. \label{eq:opr}
\end{eqnarray}
Also, the OPERA data indicate \emph{no energy
dependence}~\cite{opera}. The OPERA result has been confirmed by a
test performed using a beam with a short-bunch time-structure
allowing to measure the neutrino flight time at the single
interaction level~\cite{opera}. Interestingly, this is compatible
with the MINOS results~\cite{minos} and the earlier short-baseline
experiments~\cite{old}. From the theoretical point of view, many
groups have already studied the possible solutions or pointed out
the challenges to the OPERA anomaly~\cite{Cacciapaglia:2011ax,
AmelinoCamelia:2011dx, Giudice:2011mm, Dvali:2011mn, Mann:2011rd,
Drago:2011ua, Li:2011ue, Pfeifer:2011ve, Lingli:2011yn,
arXiv:1109.6215, Iorio:2011ay, Alexandre:2011bu, Cohen:2011hx,
GonzalezMestres:2011jc, Matone:2011jd, Ciuffoli:2011ji, Bi:2011nd,
Wang:2011sz, Cowsik:2011wv, Li:2011zm, AmelinoCamelia:2011bz, arXiv:1110.0697,
Moffat:2011ue, Faraggi:2011en, arXiv:1110.2015, Li:2011rt,
Zhao:2011sb, Chang:2011td, arXiv:1111.0093, Matone:2011fn,
Evslin:2011vq, arXiv:1111.0805, Lingli:2011kh, Mohanty:2011rm,
Li:2011ad, Ling:2011re, arXiv:1111.4994, arXiv:1111.5643,
arXiv:1111.6330, Bezrukov:2011qn}. For an early similar study, see
Ref.~\cite{Ellis:2008fc}.

The OPERA experiment would strongly imply new physics beyond the
traditional special relativity if it could be confirmed by the
future experiments. If Lorentz symmetry is broken
hardly~\cite{Colladay:1996iz, AmelinoCamelia:1997gz,
Coleman:1998ti}, {\it i.e.}, there is a preferred frame of
reference, we do have two strong constraints. The first one comes
from bremsstrahlung effects~\cite{Cohen:2011hx}. A superluminal muon
neutrino with $\delta' v_{\nu}$ given in Eq.~(\ref{eq:opr}) would
lose energy rapidly via Cherenkov-like processes on their way from
CERN to the Gran Sasso, and the most important process is $\nu_{\mu} \to
\nu_{\mu}  e^+  e^-$. Thus, the OPERA experiment would not be able
to observe muon neutrinos with energy in excess of about 12.5
GeV~\cite{Cohen:2011hx}. The second one arises from pion
decays~\cite{GonzalezMestres:2011jc, Bi:2011nd, Cowsik:2011wv}. The
superluminal muon neutrinos could not gain energy larger than about
5 GeV from the processes, $\pi^+ \to \mu^+ \nu_{\mu}$ and $\mu \to
\nu_{\mu}  e  {\bar \nu}_e$~\cite{Bi:2011nd}. However, these
constraints do not apply to other proposals which can explain the
OPERA anomaly as well~\cite{AmelinoCamelia:2011bz, Li:2011rt,
Evslin:2011vq, Lingli:2011kh, Ling:2011re,
arXiv:1111.4994, arXiv:1111.5643, arXiv:1111.6330}.

In Refs.~\cite{Cohen:2011hx,GonzalezMestres:2011jc,Bi:2011nd,
Cowsik:2011wv}, the authors considered the following simple
dispersion relation for neutrinos as a result of hard Lorentz
violation
\begin{eqnarray}
E^2_{\nu}=\vec{p}_{\nu}^{~2} + m_{\nu}^2 + \delta
\vec{p}_{\nu}^{~2} ~,~\, \label{Dispersion}
\end{eqnarray}
where $E_{\nu}$ and $\vec{p}_{\nu}$ are respectively the
neutrino energy and momentum, and  
$\delta \simeq 2 \delta' v_{\nu}$. For simplicity, we shall use
$\delta = 5 \times 10^{-5}$ in the following. The discussions on
pion decays are simple, and one can easily confirm the previous
results~\cite{GonzalezMestres:2011jc,Bi:2011nd,Cowsik:2011wv}.
However, the results for bremsstrahlung effects are a bit subtle.
From private communications with our colleagues in October 2011,
nobody seems to have succeeded in reproducing the results of Cohen
and Glashow in the approximation of four-Fermi interactions (see
also attempts in Refs.~\cite{Mohanty:2011rm, Li:2011ad}).
In addition, for neutrinos with much larger
energy $E> M_{Z}/\sqrt{\delta}$,
their decay width and energy loss
rate have not been studied as well.

In this paper, assuming the simple dispersion relation given in
Eq.~(\ref{Dispersion}), we study in detail the electron-position
emission process of superluminal neutrinos $\nu_{\mu} \to
\nu_{\mu} e^+  e^-$ for bremsstrahlung effects in the laboratory
frame, and our results can be applied to the other processes as
well. By a direct calculation, we confirm the Cohen-Glashow's
results for the low energy neutrinos. We present a general formula
for the emission process of superluminal neutrinos via a virtual
$Z$-boson exchange, which applies at both high and low neutrino
energies. Interestingly, we confirm the Cohen-Glashow's low-energy
results as well. Moreover, we systematically
 analyze the constraints on the neutrino speed from the
OPERA, Super-Kamiokande and IceCube experiments.

\section{Pair Emission Decay Width and Energy Loss Rate}

We study the pair emission process for a superluminal neutrino in
two approaches. In the first one, we work directly with the
low-energy four-Fermi interaction that is appropriate for momentum
transfer much smaller than the $Z$-boson mass $M_Z$. We recover
the results reported in Ref.~\cite{Cohen:2011hx}. We also present a new
result for the process of on-shell $Z$-boson emission. The latter
will be employed in the second approach to work out a formula that
applies also to neutrinos with general energy.

\subsection{Direct Calculations}

We start with some kinematical considerations of the pair emission
process by a superluminal neutrino,
$\nu_{\mu}(k)\to\nu_{\mu}(\ell)e^{-}(p_1)e^{+}(p_2)$, where the
quantities in the parentheses denote the four-momenta
\begin{eqnarray}
k=(E,\vec{k}\ ),\quad \ell=(E_{\nu},\vec{\ell}\ ),\quad
p_1=(E_{e^-},\vec{p}_1\ ),\quad p_2=(E_{e^+},\vec{p}_2\
)~.~\,
\label{4-momentum}
\end{eqnarray}
The neutrinos are assumed to satisfy the dispersion relation shown
in Eq.~(\ref{Dispersion}), while the electrons fulfil the usual one
\begin{eqnarray}
&E^{2}=c_{\nu}^{2}|\vec{k}|^{2}+c_{\nu}^{4}m_{\nu}^{2},\quad
E_{\nu}^{2}=c_{\nu}^{2}|\vec{\ell}|^{2}+c_{\nu}^{4}m_{\nu}^{2}\nn,&\\
&E_{e^-}^{2}=|\vec{p}_1|^{2}+m_{e}^{2},~~~~~\quad
E_{e^+}^{2}=|\vec{p}_2|^{2}+m_{e}^{2},&
\label{MassEnergyRelations}
\end{eqnarray}
where $m_{e,\nu}$ are respectively the electron and neutrino masses, $c_{\nu}
\equiv {\sqrt {1+\delta}}$ is the limiting speed of neutrinos in
units of the speed of light which we have set to be unity.

\begin{figure}[htb]
\begin{center}
\includegraphics[scale=0.9]{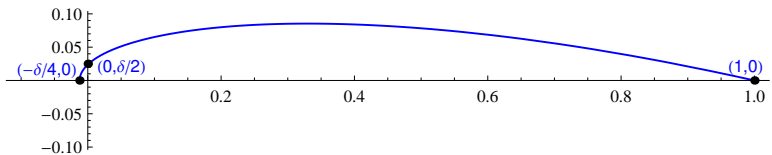}
\end{center}
\caption{ $\ell_k^\textrm{max}$ is shown in polar
coordinates as a function of $\theta_\nu$ for $\delta=0.05$. The
points $(1,0)$, $(0,\delta/2)$, and $(-\delta/4,0)$ (in rectangular
coordinates) correspond to the three kinematical configurations:
$\vec k\parallel\vec\ell$, $\vec k\perp\vec\ell$, and $-\vec
k\parallel\vec\ell$, respectively.}\label{fig:l_kmax}
\end{figure}

For the energy region of interest here, the lepton masses can be
safely neglected, so that the dispersion relations are simplified to
$E=c_{\nu}|\vec{k}|,~\textrm{and}~E_{e^-}=|\vec{p}_1|$, etc. Noting
$|\vec{p}_1+\vec{p}_2|\leq|\vec{p}_1|+|\vec{p}_2|$, one gets the
fraction of the initial neutrino energy carried away by the final
one as follows
\begin{eqnarray}
\ell_k\equiv\frac{|\vec{\ell}|}{|\vec{k}|}=\frac{E_\nu}{E}\in[0,
\ell_{k}^\textrm{max}]~,~\,
\label{lk}
\end{eqnarray}
where for a given angle $\theta_\nu$ between the moving directions
of the initial and final neutrinos we have
\begin{eqnarray}
\ell_{k}^\textrm{max}=\delta^{-1}\Delta
\big[1-\sqrt{1-\delta^2\Delta^{-2}}\big],~\Delta & = &
c_\nu^2-\cos\theta_\nu ~.~\,
\label{lkmax}
\end{eqnarray}
Fig.~\ref{fig:l_kmax} shows in polar coordinates the relation between
$\ell_k^\textrm{max}$ and $\theta_\nu$, where for illustrative
purpose we used an unrealistic value $\delta=0.05$. The figure
implies a factor of $\delta^2$ suppression from the neutrino part of
the phase space, which will contribute to the decay width.

We have done three independent calculations for the pair emission
and $Z$-boson emission processes and reached consistent results. In
what follows, we present some details about the calculations. The
amplitude for the pair emission process at momentum transfer much
smaller than $M_Z$ is
\begin{eqnarray}
{\cal M}&=&\sqrt{2}G_F\bar u(\ell)\gamma_\mu P_Lu(k)\bar
u(p_1)\gamma^\mu \big((1-2s_W^2)P_L-2s_W^2P_R\big)v(p_2)~,~\,
\end{eqnarray}
where $G_F$ is the Fermi constant, $P_{L,R}=(1\mp\gamma^5)/2$,
$s_W=\sin\theta_W$, and $u,~v$ are Dirac spinor wavefunctions for
the particles. By the way, a modified dispersion relation for neutrinos may
be formulated in Lagrangian field theory in terms of a modified
metric tensor, which in turn can affect the spin sum of neutrinos
and their effective interactions. For a careful analysis on such
subtleties or model dependencies, see Ref.~\cite{Bezrukov:2011qn}.

In this paper, we shall average over the initial neutrino spin states, 
similar to the Ref.~\cite{Cohen:2011hx}.  We are aware that the OPERA neutrinos
produced from the positively charged pion decay have a negative
helicity, and the results are roughly the same if one does not
average over the neutrino spin states~\cite{Bezrukov:2011qn}. Moreover,
 our results can be applied directly to the randomly polarized
neutrinos whose astrophysical sources, for instance, are unknown.
In short, the spin-summed and -averaged decay width and
corresponding energy loss rate are
\begin{eqnarray}
\Gamma&=&\frac{1}{2E}\int\overline{\sum_\textrm{spins}}{|{\cal
M}|^2} d\textrm{PS}_3 ~,~\,
\\
\frac{dE}{dx}&=&\frac{1}{c_\nu}\int (E_\nu-E)d\Gamma ~,~\,
\end{eqnarray}
where $d\textrm{PS}_3$ is the usual three-body phase space measure.
Neglecting the neutrino and electron masses, which is appropriate
for the OPERA experiment, the phase space of the electrons can be
easily done. We obtain
\begin{eqnarray}
\Gamma&=&\frac{8G_F^2}{96\pi}\big[(1-2s_W^2)^2+(2s_W^2)^2\big]\frac{1}{2E}J ~,~\,
\end{eqnarray}
where the phase space integral $J$ reduces to ($q=k-\ell$)
\begin{eqnarray}
J&=&\int\frac{d^3{\vec\ell}}{2E_\nu(2\pi)^3}2\big(k\cdot\ell
q^2+2k\cdot q\ell\cdot q\big) ~.~\,
\end{eqnarray}
Using the kinematics defined earlier, we obtain, e.g.,
$k^2=\delta|\vec k|^2$, and $k\cdot\ell=\Delta|\vec k||\vec\ell|$. We
find that it is easier to use the variables $\ell_k$ and $y$
instead of $|\vec\ell|$ and $\cos\theta_\nu$, where $\ell_k$ was
defined in Eq.~(\ref{lk}) and $y$ is defined by
$2\Delta=(y+y^{-1})\delta$. The integral becomes elementary
\begin{eqnarray}
J&=&\frac{|\vec k|^6}{c_\nu(2\pi)^2}\int_{y_0}^1dy
\int_0^{\ell_k^\textrm{max}}d\ell_k ~\frac{1}{2}\delta(y^{-2}-1)
\ell_k\big[3\delta\Delta\ell_k(1+\ell_k^2) -4\Delta^2\ell_k^2
-2\delta^2\ell_k^2\big]~,
\end{eqnarray}
where $y_0=(c_\nu-1)/(c_\nu+1)$. Working out the integral we get
\begin{eqnarray}
\Gamma&=&\frac{G_F^2E^5\delta^3}{192\pi^3c_\nu^7}
\big[(1-2s_W^2)^2+(2s_W^2)^2\big]
\bigg\{\frac{1}{7}+\frac{y_0^7}{140}-\frac{y_0^5}{20}
+\frac{3y_0^3}{20}-\frac{y_0}{4}\bigg\} ~.~\,
\end{eqnarray}
This result is exact with the only approximation being
$m_{e,\nu}=0$. For the energy loss rate, one multiplies the
integrand of $J$ by $-c_\nu^{-1}E(1-\ell_k)$ and obtains
\begin{eqnarray}
\frac{dE}{dx}&=&-\frac{G_F^2E^6\delta^3}{192\pi^3c_\nu^8}
\big[(1-2s_W^2)^2+(2s_W^2)^2\big]
\nonumber
\\
&\times&\bigg\{\frac{25}{224}-\bigg[\frac{1}{160}y^8_0-\frac{1}{140}y^7_0
-\frac{3}{80}y^6_0+\frac{1}{20}y^5_0+\frac{7}{80}y^4_0-\frac{3}{20}y^3_0
-\frac{7}{80}y^2_0+\frac{1}{4}y_0\bigg]\bigg\} ~.~\,
\end{eqnarray}
Putting $s_W^2=1/4$ and keeping only the leading term in
$\delta\ll 1$ we recover the Cohen-Glashow results
\cite{Cohen:2011hx}
\begin{eqnarray}
\Gamma&\approx&\frac{1}{14}\frac{G_F^2E^5\delta^3}{192\pi^3} ~,~\,
\label{CG1}
\\
\frac{dE}{dx}&\approx&-\frac{25}{448}\frac{G_F^2E^6\delta^3}{192\pi^3} ~.~\,
\label{CG2}
\end{eqnarray}

For a superluminal neutrino with high enough energy, the process
$\nu(k)\to\nu(\ell)Z(p)$ can also take place. Ignoring again the
neutrino mass, the spin-summed and -averaged amplitude squared is
\begin{eqnarray}
\overline{\sum_\textrm{spins}}|{\cal M}|^2&=&\sqrt{2}G_F
\big[k\cdot\ell M_Z^2+2k\cdot p\ell\cdot p\big] ~.~\,
\end{eqnarray}
The energy-momentum conservation gives
\begin{eqnarray}
z^2-2z\Delta\delta^{-1}+1-\ell_Z^2=0 ~,~\,
\end{eqnarray}
where $z=|\vec\ell|/|\vec k|$ is similar to $\ell_k$ in the pair
emission process, $\Delta$ was defined in Eq.~(\ref{lkmax}), and
$\ell_Z=(c_\nu M_Z)/(E\sqrt{\delta})$. The above as an equation of
$z$ for given $\cos\theta_\nu\in[-1,1]$ has always two real
solutions for $c_\nu>1$, but one of them being larger than unity
is non-physical. Requiring the other solution to be in the
physical region yields the threshold condition, $\ell_Z<1$, {\it
i.e.}, $E>c_\nu M_Z/\sqrt{\delta}$. Similarly to the case of pair
emission, it is simpler to work in phase space with $z$ instead of
$\Delta$ or $\cos\theta_\nu$,
$2\Delta=[z+(1-\ell_Z^2)z^{-1}]\delta$, and the interval for
$\cos\theta_\nu$ translates into $z\in[z_-,z_+]$ with
\begin{eqnarray}
z_-&=&(2+\delta)\delta^{-1}\Big[1-
\sqrt{1-(2+\delta)^{-2}\delta^2(1-\ell_Z^2)}\Big]~, \nonumber
\\
z_+&=&1-\ell_Z~.
\end{eqnarray}
Using the new variables, we have, e.g.,
$2k\cdot\ell=c_\nu^{-2}E^2(1+z^2-\ell_Z^2)\delta$, and $2p\cdot
k=c_\nu^{-2}E^2(1-z^2+\ell_Z^2)\delta$. The decay width and the
neutrino energy loss rate are found to be
\begin{eqnarray}
\Gamma&=&\frac{1}{16\pi}\frac{G_F}{\sqrt{2}}\frac{E^3}{c_\nu^7}I_\Gamma,
\\
\frac{dE}{dx}&=&-\frac{1}{16\pi}\frac{G_F}{\sqrt{2}}\frac{E^4}{c_\nu^8}I_R,
\end{eqnarray}
where the residual phase space integrals are elementary
\begin{eqnarray}
I_\Gamma&=&c_\nu^2\delta^2\int_{z_-}^{z_+}dz~\big[\ell_Z^2(1+z^2)
+(1-z^2)^2-2\ell_Z^4\big]~,
\\
I_R&=&c_\nu^2\delta^2\int_{z_-}^{z_+}dz~\big[\ell_Z^2(1+z^2)
+(1-z^2)^2-2\ell_Z^4\big](1-z) ~.~\,
\end{eqnarray}
Again the above results only assume $m_{e,\nu}=0$ and are exact in
parameters $\ell_Z$ and $\delta$. Since $\delta\ll 1$, we can
expand in $\delta$ while holding $\ell_Z<1$ fixed and obtain the
results in the leading order of $\delta$
\begin{eqnarray}
\Gamma&\approx&\frac{G_FE^3\delta^2}{120\sqrt{2}\pi}
\big(4+10\ell_Z^2-25\ell_Z^3+11\ell_Z^5\big)~,%
\label{eq_Z_rate}
\\
\frac{dE}{dx}&\approx&-\frac{G_FE^4\delta^2}{960\sqrt{2}\pi}
\big(22+35\ell_Z^2-180\ell_Z^4+88\ell_Z^5+35\ell_Z^6\big) ~,~\,
\label{eq_Z_loss}
\end{eqnarray}
where now $\ell_Z\approx M_Z/(E\sqrt{\delta})$.

\subsection{Virtual $Z$ Approach }

In the previous subsection we computed the electron-positron
emission in the low energy limit and the emission of a physical
$Z$ boson at sufficiently high energies. In the following we
present our general results for the pair emission via a virtual
$Z$ exchange, $\nu(k)\to\nu(\ell)Z^*(q)\to\nu(\ell)f(p_1)\bar
f(p_2)$, which can be applied at any initial neutrino energy $E$.
One can do so purely numerically of course, but we would like to
accomplish this in such a way that both low and high energy limits
can be readily identified. After some manipulations of phase space
and appropriate reorganization of amplitudes for the subprocesses,
we obtain upon neglecting the neutrino masses
\begin{eqnarray} \Gamma&=&\frac{1}{\pi}\int_0^{\delta E^2}dm_*^2
\frac{m_*\Gamma_f(m_*)}{(m_*^2-M_{Z}^2)^2+\Gamma_Z^2M_{Z}^2}
\Gamma_i(m_*),%
\label{eq_general_rate}
\\
\frac{d E}{d x}&=&\frac{1}{\pi}\int_0^{\delta E^2}dm_*^2
\frac{m_*\Gamma_f(m_*)}
{(m_*^2-M_{Z}^2)^2+\Gamma_Z^2M_{Z}^2} \frac{d E_i(m_*)}{d x} ~,~\,
\label{eq_general_loss}
\end{eqnarray}
where {\bf $m_*=\sqrt{q^2}$}, $\Gamma_Z$ is the usual total decay
width of the $Z$ boson, $\Gamma_i(m_*)$ is the decay width of the
initial superluminal neutrino into a virtual $Z$ boson with an
effective mass $m_*$, and $dE_i(m_*)/dx$ is the corresponding
energy loss rate. These functions are obtained from
Eqs.~(\ref{eq_Z_rate}) and (\ref{eq_Z_loss}) by replacing $M_Z$
with $m_*$ everywhere, {\it i.e.},
$\ell_Z\to\ell_{Z^*}=\ell_Zm_*/M_Z$ and $G_F\to
G_F^*=G_FM_Z^2/m_*^2$. Similarly, $\Gamma_f(m_*)$ is the decay
width of a virtual $Z$ into a pair of fermions. The above results
can be used to individual channels that are kinematically allowed,
although in numerical analysis we will show the results summing
over all possible channels.

The above Eqs.~(\ref{eq_general_rate}) and (\ref{eq_general_loss})
serve as a useful function interpolating between the low and high
energy limits. At energies much lower than $M_Z$ but still above
the threshold for the electron-positron pair, the denominator is
approximately equal to $M_Z^4$, and the Cohen-Glashow's results in
Eqs.~(\ref{CG1}) and (\ref{CG2}) are recovered. In the high energy
limit $E\gg M_Z/\sqrt{\delta}$, the integrals can be worked out in
the narrow width approximation (NWA) (see, e.g., Appendix B in
Ref. \cite{Han:2005mu}). When summing over all decay channels of
the  virtual $Z$ boson at sufficiently high energy, the
results in Eqs.~(\ref{eq_Z_rate}) and (\ref{eq_Z_loss}) for the
physical $Z$-boson emission are recovered as well.

\section{Numerical Analyses for Neutrino Velocities in Various Experiments}

\subsection{General Discussion}

\begin{figure}[htb]
\begin{center}
\includegraphics[scale=0.55]{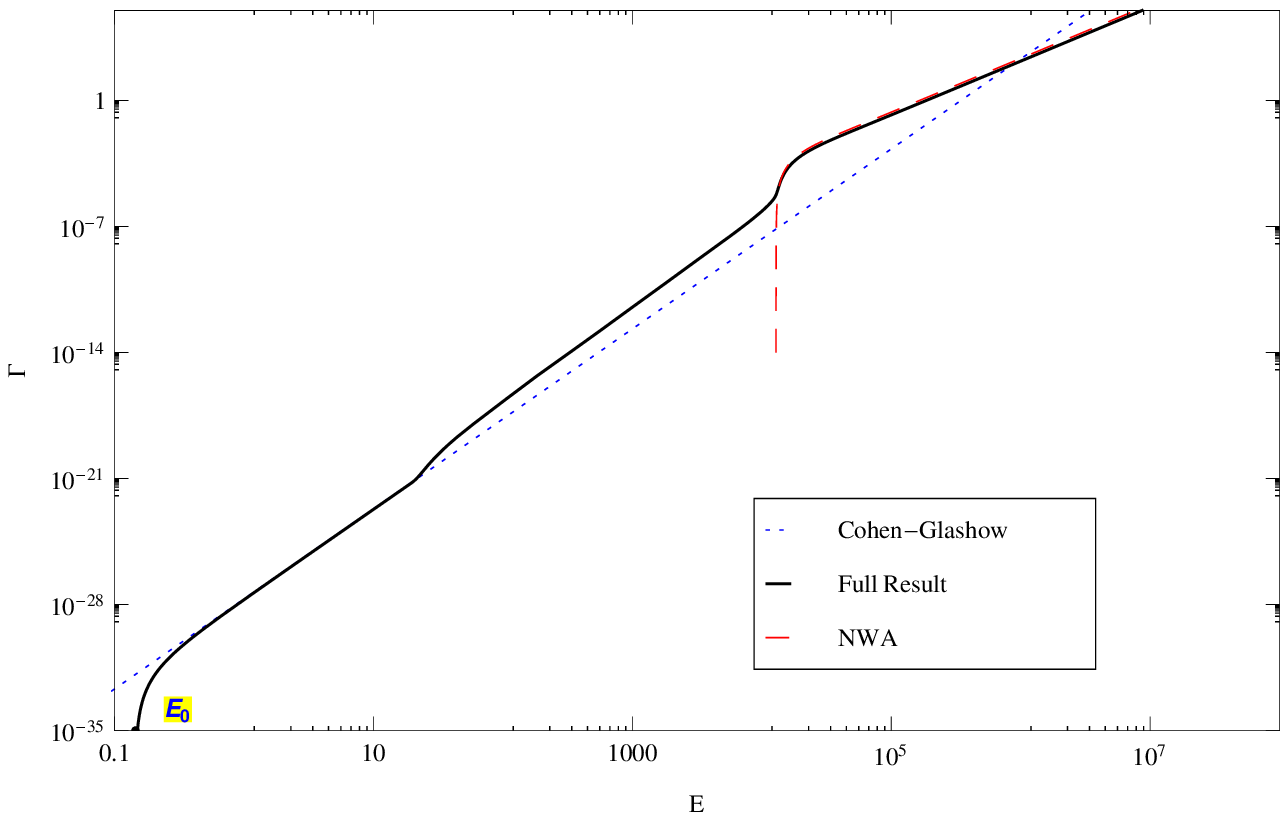}
\includegraphics[scale=0.55]{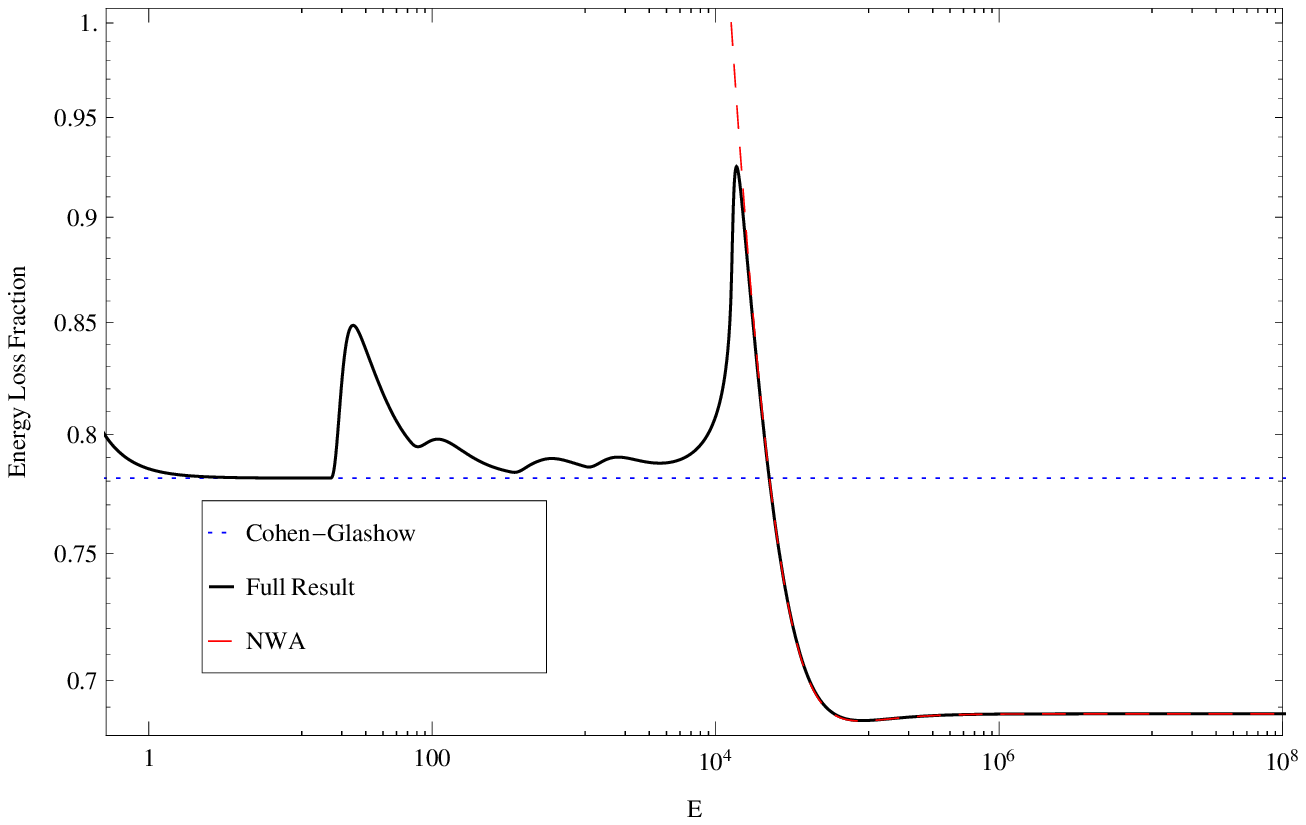}\\
\end{center}
\caption{ Decay width $\Gamma$ (in GeV, left panel) and
energy loss fraction $-(dE/dx)/(E\Gamma)$ (right panel) of a
superluminal neutrino are shown as a function of the initial
energy $E$ (in GeV), using the three results: full result
(solid-black curve), Cohen-Glashow's result (dotted-blue) and NWA
(dashed-red). Here $\delta=5\times 10^{-5}$, and
$E_0=2m_e/\sqrt{\delta}$ is the
threshold for the electron-positron emission.}%
\label{radiation_rate}
\end{figure}

We first compare in Fig.~\ref{radiation_rate} numerical results
based on Eqs.~(\ref{eq_general_rate}) and (\ref{eq_general_loss}) with
those in the low-energy limit in Ref.~\cite{Cohen:2011hx} and in
the high energy limit using NWA, Eqs.~(\ref{eq_Z_rate}) and
(\ref{eq_Z_loss}). To avoid misunderstanding, we note that while
the low-energy result involves only the electron-positron
emission, the other two include all possible channels that are
kinematically allowed. Due to the opening of more and more
channels one sees a continuous enhancement of the decay width (in
the left panel) and energy loss rate (in the right panel). At very high
energies the NWA result essentially coincides with the full
calculation.

The difference to the low-energy limit can be more clearly seen in
the right panel of Fig.~\ref{radiation_rate} for the  energy
loss fraction. The mean fractional energy loss is about 0.79 for
most energy scales until the $Z$ threshold (main peak). At the
threshold, the process is dominated by the production of an
on-shell $Z$ which takes away most of the initial neutrino energy.
Beyond that point, there will be more energy left for the final
neutrino. The secondary peak at about 30 GeV is due to the
setting-in of the $u\bar{u}$, $d\bar{d}$, $\mu^+\mu^-$, and
$s\bar{s}$ final states. It is worthy noticing that the energy
loss rate decreases down to approximately 0.69 in the high energy
limit where the low-energy approximation is not applicable.

\subsection{The OPERA Experiment}

Based on the energy loss rate at low energies in Eq.~(\ref{CG2})
Cohen and Glashow defined a terminal energy $E_T$ for a
superluminal neutrino after travelling a distance of $L$~\cite{Cohen:2011hx}
\begin{eqnarray}
 E_{T}^{-5}=\frac{125}{448}\frac{G_{F}^2L\delta^3}{192\pi^3}~,~\,
\end{eqnarray}
meaning that a neutrino with energy less than $E_T$ would not have
travelled such a long distance. For the OPERA neutrinos with
$L=730~\textrm{km}$ and $\delta=5\times 10^{-5}$, they got
$E_T=12.5~\textrm{GeV}$. We observe from Eq.~(\ref{CG1}) that a
superluminal neutrino with energy $E_T$ has a decay length,
$c/\Gamma$, that is of the same order as the travel distance $L$.
We thus propose to use a different statistical measure, the
survival probability
\begin{eqnarray}
P=e^{-L\Gamma}~,~\,
\end{eqnarray}
 to decide whether a
neutrino is observable or not at the OPERA detector.

\begin{figure}[htb]
\begin{center}
\includegraphics[scale=0.56]{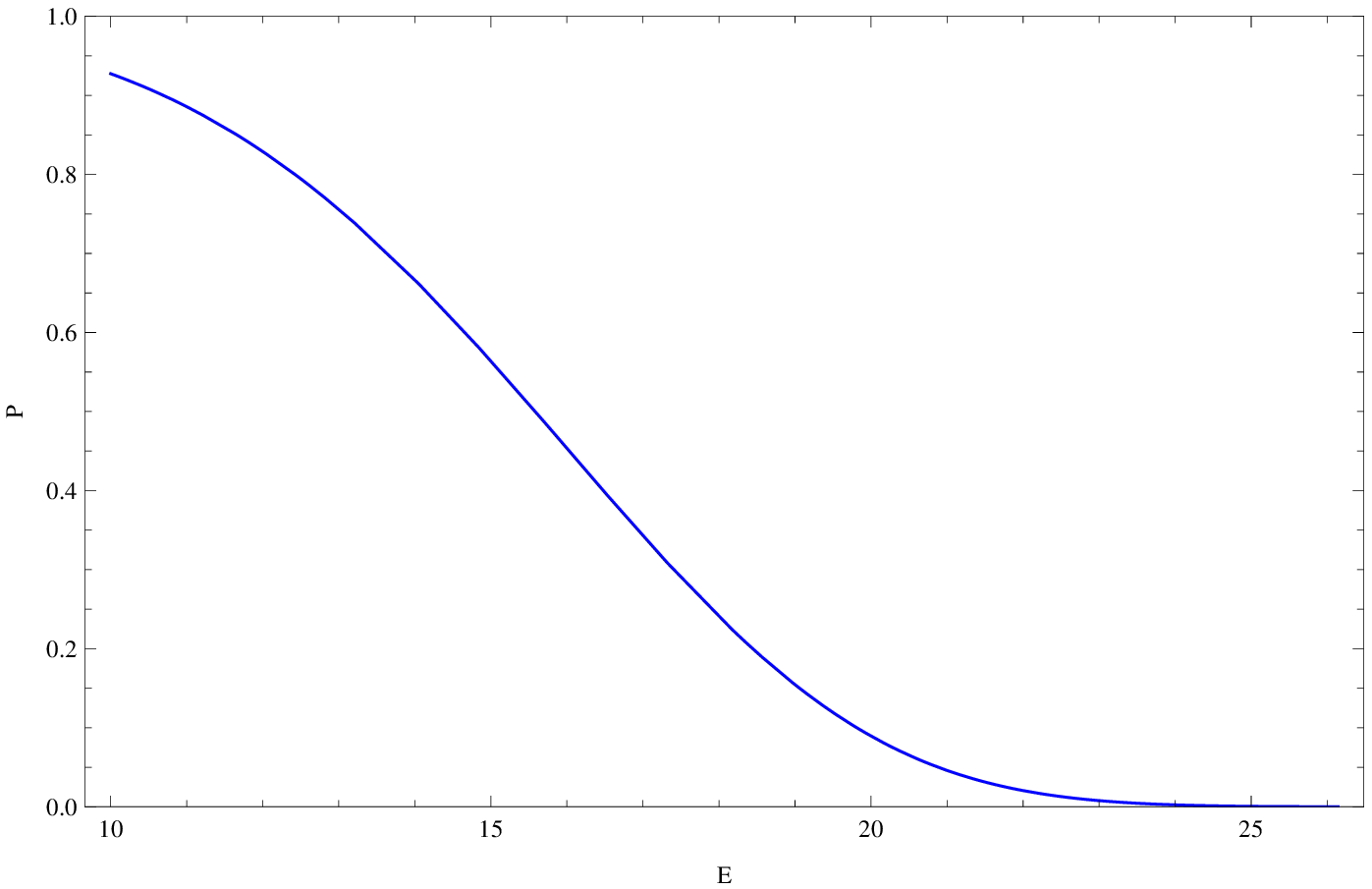}
\includegraphics[scale=0.56]{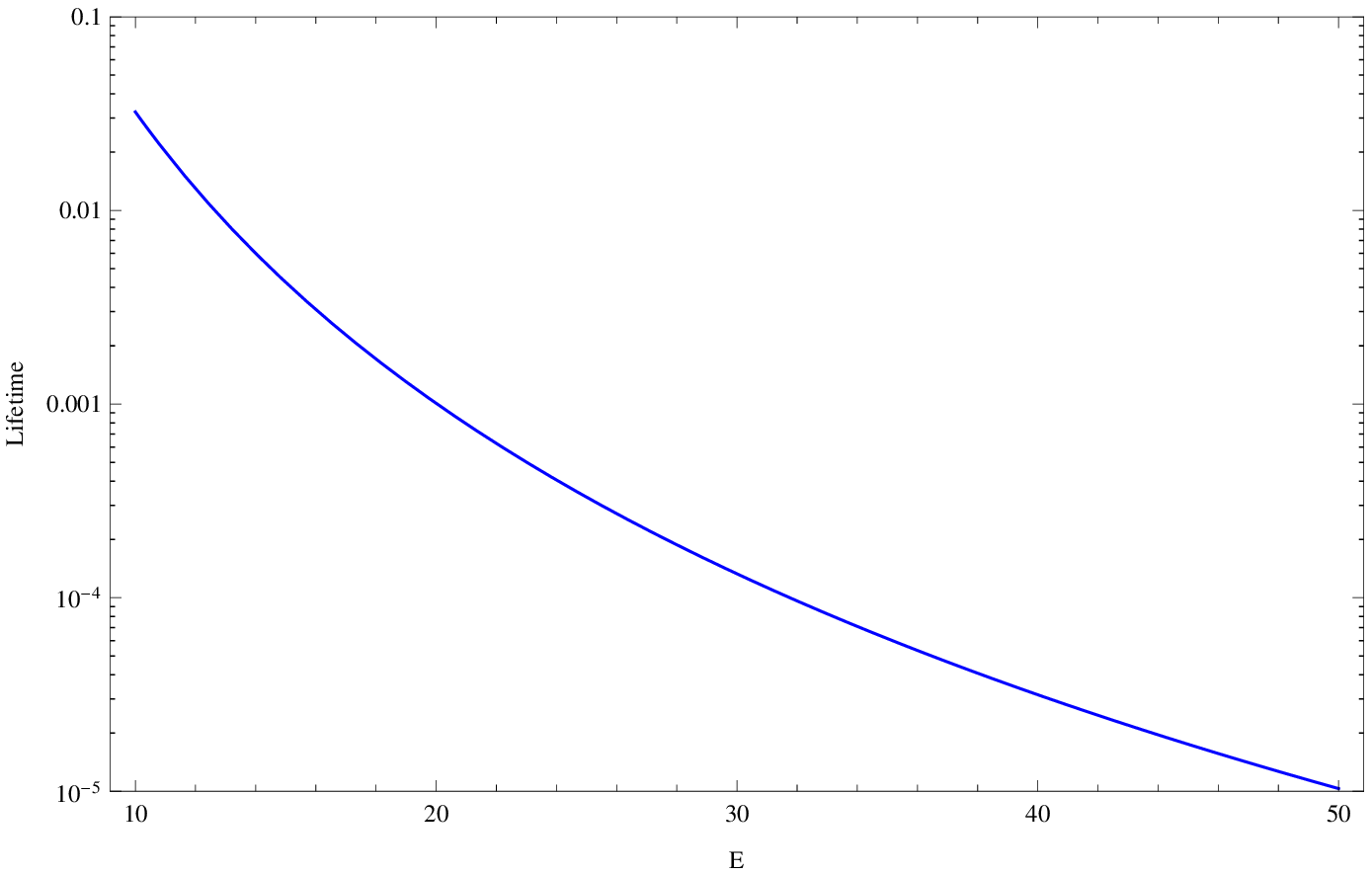}
\end{center}
\caption{ Survival probability $P$ (left panel) and
lifetime (in seconds, right panel) as a function of energy $E$ (in
GeV) in the OPERA experiment with $\delta=5\times10^{-5}$ and
$L=730~\textrm{km}$.}%
\label{Undecay_Rate}
\end{figure}

To get some feeling of numbers, we notice that a superluminal neutrino of
mean energy $\sim 17.5~\textrm{GeV}$ has a lifetime of $1.9\times
10^{-3}~\s$, which should be compared with its travel time of
$2.4\times 10^{-3}~\s$ if it happens to have not yet decayed during
the trip. The survival probability can tell in a statistically
better way how likely such a neutrino can indeed be observable. From
the left panel in Fig.~\ref{Undecay_Rate} one can see that a
neutrino has a survival probability of $0.79,~0.29,~0.09$
respectively when it carries a terminal energy $12.5~\textrm{GeV}$,
mean energy $17.5~\textrm{GeV}$, and an
energy of $20~\textrm{GeV}$. In particular, a neutrino with an
initial energy of order $E_T$ has a good chance to arrive at the
OPERA detector. The right panel of Fig.~\ref{Undecay_Rate} shows
that the neutrino lifetime drops drastically as its energy
increases. In that case the terminal energy serves as an appropriate
measure on its observability.

\subsection{The Super-Kamiokande and IceCube Experiments}

Besides OPERA, the Super-Kamiokande and IceCube experiments also
reported observation of high energy neutrinos that travel a long
distance. The Super-Kamiokande detected atmospheric neutrinos that
traverse a distance of $10^4~\textrm{km}$ through the Earth
(upward-going in the detector) over an energy range extending from 1
GeV to 1 TeV \cite{Ashie:2005ik,Swanson:2006gm,Desai:2007ra}. The
IceCube collaboration observed upward-going showers with
reconstructed shower energies above 16 TeV~\cite{Abbasi:2011ui},
with a baseline length estimated to be at least 500 km. Moreover, it
reported the upward-going neutrinos from 100 GeV to 400 TeV
\cite{arXiv:1010.3980}. For the latter we choose a rough travel
distance about $10^4$ km for the neutrinos within a zenith angle of
$124^\textrm{o}-180^\textrm{o}$.

\begin{figure}[htb]
\begin{center}
\includegraphics[scale=1.1]{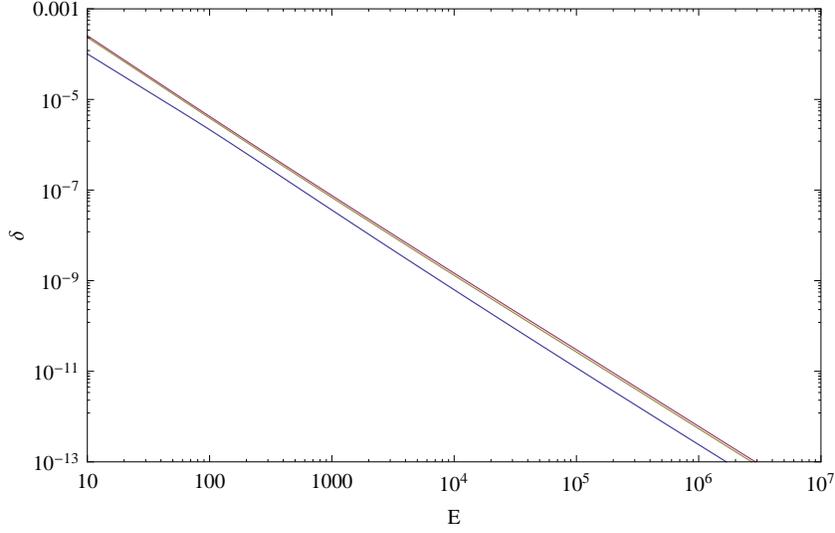}
\end{center}
\caption{Constraints on $\delta$ as a function of
energy $E$ (in GeV) for three baseline lengths
$L=500,~730,~\textrm{and} ~10^4~\textrm{km}$
from the top to bottom lines.}%
\label{Constraint_on_Delta}
\end{figure}

We perform an analysis of constraints on the speed deviation
$\delta$ of neutrinos from that of light in those two experiments
using our general result in Eq.~(\ref{eq_general_rate}). We require
that the neutrinos detected in Super-Kamiokande and IceCube
experiments have travelled a distance $L$ that is smaller than $n$
times their decay length $c_\nu/\Gamma$ at the given energy $E$,
{\it i.e.}, $L<n c_\nu/\Gamma$. Since $\Gamma$ depends roughly on
positive powers of $E$ and $\delta$ (e.g., $E^5\delta^3$ at low
energy), this gives an upper bound on $\delta$ as a function of $E$.
The bound is not sensitive to $n$ as long as $n$ is not very
different from unity, and a larger $n$ yields a more conservative
bound on $\delta$ using the same data. Our results assuming $n=10$
are shown in Fig.~\ref{Constraint_on_Delta} for three baseline lengths
$L=500,~730,~\textrm{and}~10^4~\textrm{km}$. For instance, for
$L=500~\textrm{km}$ the observations of neutrinos with energy in excess
of $16~\textrm{TeV}$ imply $\delta<6.6\times10^{-10}$, while
for $L=10^4~\textrm{km}$ the
observations of neutrinos with $E\ge 400~\textrm{TeV}$
 require $\delta<7.8 \times 10^{-12}$. These
bounds are indeed very stringent.

\begin{figure}[htb]
\begin{center}
\includegraphics[scale=0.7]{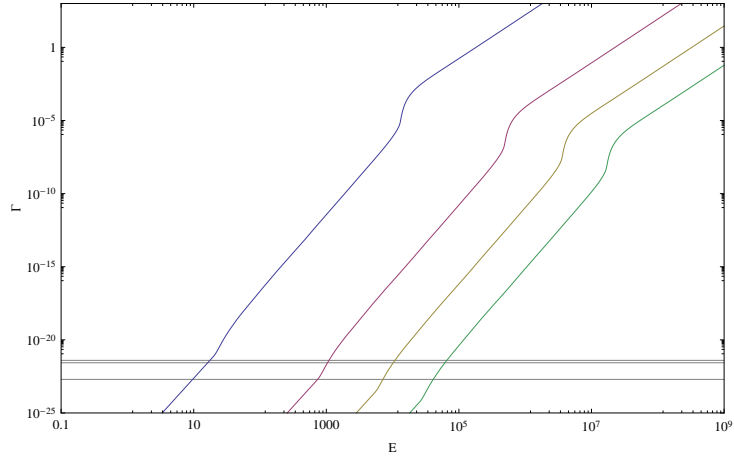}
\end{center}
\caption{ Decay width $\Gamma$ (in GeV) as a function
of energy $E$ (in GeV) for various $\delta=5\times10^{-5}$,
$3.6\times10^{-8}$,  $6.6\times10^{-10}$, and $3.0\times10^{-11}$ from the
left to right curves. The horizontal lines indicate the decay length
equal to the baseline lengths $L=500,~730,~\textrm{and}~10^4~\textrm{km}$ from top to
bottom.}%
\label{Constraints_on_speed_departure}
\end{figure}

As both analytical results and Fig.~\ref{Constraint_on_Delta}
indicate, the bounds on $\delta$ are not sensitive to the baseline length
$L$ at a given neutrino energy $E$ but are rather sensitive to $E$.
The simple straight-line behavior implies that the extreme high
energy regime has not yet been touched in those observations. To get
some idea of the regime, we plot in
Fig.~\ref{Constraints_on_speed_departure} the decay width $\Gamma$ as a
function of $E$ for four values of $\delta$ that correspond
respectively to the bounds obtained from the experiments OPERA,
Super-Kamiokande with $(L,E)=(10^4~\textrm{km},~1~\textrm{TeV})$, and
IceCube with
$(L,E)=(500~\textrm{km},~16~\textrm{TeV})$ and $(500~\textrm{km},~100~\textrm{TeV})$.
These experiments available on the Earth constrain the $\delta$
parameter to such tiny values that it is hard to reach the high
energy regime with $E>M_Z/\sqrt{\delta}$.

\section{Conclusions }

We revisited the constraints on superluminal neutrino velocities
by assuming a minimally modified dispersion relation. We obtained
the general decay width and energy loss rate of a superluminal
neutrino with high or low energy from the bremsstrahlung process
via both the direct calculation and virtual $Z$ approaches in the
laboratory frame. The analytical results by Cohen and Glashow were
confirmed in the low energy limit. We used a different measure to
assess whether a neutrino is observable or not in the OPERA
experiment. We presented new results on the power law for the
bremsstrahlung process in the high energy limit. Using our general
results, we performed systematical analyses on the constraints
arising from the Super-Kamiokande and IceCube experiments.

\section*{Acknowledgments}

We would like to thank Xiaojun Bi, Jarah Evslin, Miao Li, Pengfei
Yin and Xinmin Zhang for helpful discussions. This research was
supported in part by the Natural Science Foundation of China under
grant numbers 10821504 and 11075194 (YH, TL and YQ), and 10975078
and 11025525 (YL), and by the DOE grant DE-FG03-95-Er-40917 (TL and
DVN).

\end{document}